\documentclass{article}
\usepackage{epsfig}
\usepackage{psfig}
\usepackage{latexsym}

\begin{document}

\begin{flushright}
   July 2004 \\
\end{flushright}

\medskip

\bigskip

\begin{center}
{\LARGE B and B}$_{S}${\LARGE \ decay constants from moments of}

{\LARGE Finite Energy Sum Rules in QCD\footnote{{\LARGE {\footnotesize %
Supported by MCYT-FEDER under contract FPA2002-00612.}}}}\vspace{2cm}

{\large F. Bod\'{i}-Esteban,} {\large J. Bordes} {\large and J.
Pe\~{n}arrocha}$\bigskip $

{\normalsize Departamento de F\'{\i}sica Te\'{o}rica-IFIC, Universitat de
Valencia}

{\normalsize E-46100 Burjassot-Valencia, Spain}

\bigskip

\bigskip

\bigskip

\textbf{Abstract\medskip }
\end{center}

\begin{quote}
We use an appropriate combination of moments of finite energy sum rules in
QCD in order to compute the $B_{q}$-meson decay constants $f_{B}$ and $%
f_{B_{s}}$.We perform the calculation using a two-loop computation of the imaginary part of the
pseudoscalar two point function in terms of the running bottom quark mass.
The results are stable with the so called QCD duality threshold and they are
in agreement with the estimates obtained from Borel transform QCD sum rules
and lattice computations. 

\bigskip
\end{quote}

PACS: 12.38.Bx, 12.38.Lg. {\newpage }

\section{Introduction}

Since the pioneering work of Shifman, Vainshtein and Zacharov \cite{SHIFMAN},
Laplace sum rules have been successfully applied to calculate all sort of
parameters of the hadronic spectrum. The main advantage of this type of sum
rules is that a Borel transform applied to the correlation function enhances
the contribution of the low lying resonances of the hadronic spectrum, 
which properties are to be determined. On the other hand, it reinforces the
convergence of the QCD asymptotic calculation in the high energy domain. The
price to pay with this method is the appearance of the so-called Borel
parameter which has to be fixed by stability arguments. Other
particular sum rules based on Hilbert transform, inverse moments.., have
also been used to suitably deal with other particular problems.

In this note we introduce a method based on positive moments of QCD finite
energy sum rules. Traditionally, this type of sum rules have the disadvantage
of reducing the contribution of the low energy part of the hadronic
spectrum, whereas they enhance the QCD high energy region. Although they
are easy to handle, one needs to fix a QCD duality threshold, where
theoretical calculations are accurate enough and at the same time the low
energy region admits a suitable hadronic parametrization. Nevertheless, it
is not always easy to fix the value of this duality threshold in the sense
that the results have to be independent of this value. The
method we propose is to combine different moments of finite energy sum rules
in order to get a polynomial weight for the correlation function such that
its contribution becomes negligible when integrated from the continuous
physical threshold until the QCD duality threshold. In this way, the low
lying resonance region is enhanced in the hadronic spectrum and, on the
other hand, the role of the QCD duality threshold becomes less relevant on
the final results. We will be more explicit on this in the next section.

Here we use our method to evaluate the decay constants of the lightest
pseudoscalar bottom mesons ($f_{B}$ and $f_{B_{s}}$), which parametrize
the $B_{q}$-meson matrix elements of the pseudoscalar current: 
\[
<\Omega |\,(M_{b}+m_{q})\,\overline{q}\,i\,\gamma _{5}\,b)(0)\,|B_{q}>\,\,
=\,\,f_{B_{q}}\,M_{B_{q}}^{2}. 
\]

These decay constants have received recently a lot of attention since they
enter in hadronic matrix elements of $B-\overline{B}$ mixing,\ and its
accurate evaluation would facilitate a better determination of the $B_{B}$
mixing factor from recent experiments carried out in the B-factories. They
also appear in the leptonic $B$ decay widths and its knowledge could provide
a good determination of the $\left| V_{qb}\right| $ matrix element in future
experiments. Calculations of these decay constants have been performed since
the eighties, with results in the range of $f_{B}=160-210$ $\mathrm{MeV}$
and $f_{B_{s}}/f_{B}=1.09-1.22$ from Borel transform techniques \cite
{REINDERS,JAMINLANGE,NARISON,DOMINGUEZPAVER}. Computations in lattice QCD give also results
in a wide range: $f_{B}=161-218\,\mathrm{MeV}$ and $%
f_{B_{s}}/f_{B}=1.11-1.16$ \cite{APE,UKQCD} (for a review and a
collection of the results, see \cite{COLANGELO}). As we see, in these 
range of values there is still room for improvement.

The plan of this note is the following: in the next section we briefly
review the theoretical method proposed, in the third section we discuss the
theoretical and experimental inputs used in the calculation and in the
fourth one we present our results for the decay constants with a discussion
of the errors. We finish the paper giving the conclusions.

\bigskip

\section{The method}

The two point function relevant to our problem is: 
\[
\Pi (s=q^{2})\,=\,i\int \,dx\,\, e^{iqx}<\Omega
|T(j_{5}(x)j_{5}(0))\,|\Omega >, 
\]
where $<\Omega \,|$ is the physical vacuum and the current $j_{5}(x)$ is the
divergence of the axial-vector current: 
\[
j_{5}(x)\,=\,(M_{Q}+m_{q}):\overline{q}(x)\,i\gamma _{5}\,Q(x): 
\]
$M_{Q}$ is the mass of the heavy quark $Q(x)$ which will the bottom
quark in our case, whereas $m_{q}$ stands for the light quark mass, 
up, down or strange. In order to write down the sum rules relevant to our
calculation, we apply Cauchy's theorem to the two point correlation function 
$\Pi (s)$, weighted with a polynomial $P(s)$ as indicated in the
following: 
\begin{equation}
\frac{1}{2\pi i}\oint_{\Gamma }s^{i}P(s)\Pi (s)\, \, ds \, = \, 0,
\label{CAUCHY} 
\end{equation}
(the power $s^{i}$, with $i \geq 0$, is introduced here for convenience, as it will become
apparent in the following).

The integration path $\Gamma $\ is extended along a circle of radius $\left|
s\right| =s_{0}$, and along both sides of the physical cut starting at the
physical threshold $s_{\rm ph.}$, i. e. running in the interval $s\,\in \,\left[
s_{\rm ph.},s_{0}\right] $. Neither the polynomial $P(s)$ nor the power of the
integration variable change the analytical properties of $\Pi (s)$, so that
we obtain the following sum rule: 
\begin{equation}
\frac{1}{\pi }\int_{s_{\rm ph.}}^{s_{0}}s^{i}\,P(s)\,\mathrm{Im}\,\Pi (s)\,ds \, = \, - \, %
\frac{1}{2\pi i}\oint_{\left| s\right| =s_{0}}s^{i}\,P(s)\,\Pi (s)\,ds
\label{SR}
\end{equation}

On the left hand side of this equation, we enter the experimental
information of $\mathrm{Im}\,\Pi (s)$, starting from the physical threshold $%
s_{\rm ph.}$ till the integration radius $s_{0}$, whereas, on the right hand
side, we consider the asymptotic QCD theoretical calculations to be plugged into
the integration contour of radius $s_{0}$. Therefore, this radius makes the
compromise where the contribution of the hadronic spectrum can be
approximated by the QCD calculation. This is our QCD duality threshold that
we refer to in the previous section.

The asymptotic expansion of QCD ($\Pi ^{\mathrm{QCD}}(s)$) can be split in
two parts, including the perturbative and non-perturbative terms, as
follows: 
\begin{equation}
\Pi ^{\mathrm{QCD}}(s)\,=\,\Pi ^{\mathrm{pert.}}(s)\,+\,\Pi ^{\mathrm{%
nonpert.}}(s),  \label{2QCD}
\end{equation}

At this stage, we consider that $\Pi ^{\mathrm{pert.}}(s)$ is an analytic
function of $s$, with a real cut starting at $s_{\rm QCD}=\left(
M_{b}+m_{q}\right) ^{2}$, therefore, we can use again the Cauchy's theorem
to convert the integration of $\Pi ^{\mathrm{pert.}}(s)$ along the circle $%
\left| s\right| =s_{0}$ into an integration of the corresponding absorptive
part along the QCD cut, 
\begin{eqnarray}
&&\frac{1}{\pi }\int_{s_{\rm ph.}}^{s_{0}}s^{i}P(s)\mathrm{Im}\,\Pi (s) \,\, ds \, = \nonumber \\
&=&\frac{1}{\pi }\int_{\left( M_{b}+m_{q}\right) ^{2}}^{s_{0}}s^{i}P(s)%
\mathrm{Im}\Pi ^{\rm pert.}(s)\,\, ds-\frac{1}{2\pi i}\oint_{\left|
s\right| =s_{0}}s^{i}P(s)\Pi ^{\mathrm{nonpert.}}(s) \, \, ds \nonumber \\
\label{EQU}
\end{eqnarray}

On the right hand side of equation (\ref{EQU}) we take for the perturbative spectral function 
the exact two-loop QCD calculation \cite{REINDERS}

\begin{eqnarray}
&& \frac{1}{\pi }\mathrm{Im}\Pi ^{\rm pert.}(s)  \, = \nonumber \\
&& = \, \frac{3}{8\pi ^{2}%
}(M_{b}+m_{q})^{2}s\left( 1-\frac{M_{b}^{2}}{s}\right) ^{2}\left\{ 1+\frac{%
\alpha _{s}\left( \mu \right) }{\pi }\frac{2}{3}\left[ 4 {\rm dilog}\left( 
\frac{M_{b}^{2}}{s}\right) \right. \right. \nonumber \\
&&+2\ln \left( \frac{M_{b}^{2}}{s}\right) \ln \left( 1-\frac{M_{b}^{2}}{s}%
\right) -\left( 5-\frac{2M_{b}^{2}}{s}\right) \ln \left( 1-\frac{M_{b}^{2}}{s%
}\right) \nonumber \\
&&+\left[ \left( 1-2\frac{M_{b}^{2}}{s}\right) \left( 3-\frac{M_{b}^{2}}{s}%
\right) \ln \left( \frac{M_{b}^{2}}{s}\right) +\frac{17}{2}-\frac{33M_{b}^{2}%
}{2s}\right. \nonumber \\
&&\left. \left. \left. -3\left( 1-\frac{3M_{b}^{2}}{s}\right) \ln \left( 
\frac{M_{b}^{2}}{\mu ^{2}}\right) \right] \left( 1-\frac{M_{b}^{2}}{s}
\right) ^{-1}\right] \right\}
\label{PERT}
\end{eqnarray}
where $M_{b}=M_{b}\left( \mu \right) $ is the running mass of the $b$\ quark
in the $\overline{MS}$ scheme. It is known that the expansion in terms of
the running mass converges much faster, in a wide range of the
renormalization scale, than the one in terms of the bottom pole mass, as
notice, for instance, in \cite{JAMINLANGE}.

For the non perturbative part $\Pi ^{\mathrm{nonpert.}}(s)$ we take the
contribution coming from the vacuum expectation values of non perturbative
operators up to dimension six \cite{ALIEV,REINDERS}: 
\begin{eqnarray}
&& \Pi ^{\mathrm{nonpert.}} (s) \,  = \, \frac{M_{\rm pole}^{2}}{s-M_{\rm pole}^{2}}\left[
M_{\rm pole}<\overline{q}q>-\frac{1}{12}<\frac{\alpha _{s}}{\pi }G^{2}>\right] -
\nonumber \\
&-&\frac{1}{2}M_{\rm pole}^{3}\left[ \frac{1}{(s-M_{\rm pole}^{2})^{2}}+\frac{%
M_{\rm pole}^{2}}{(s-M_{\rm pole}^{2})^{3}}\right] <\overline{q}\sigma Gq>- 
\nonumber \\
&-&\frac{8}{27}\pi M_{\rm pole}^{2}\left[ \frac{2}{(s-M_{\rm pole}^{2})^{2}}+\frac{%
M_{\rm pole}^{2}}{(s-M_{\rm pole}^{2})^{3}}-\frac{M_{\rm pole}^{4}}{(s-M_{\rm pole}^{2})^{4}}%
\right] \alpha _{s}<\overline{q}q>^{2}.  \nonumber \\
\label{NOPERT}
\end{eqnarray}

The contour integration of the non perturbative part is easily done by means
of the residues theorem.

Again, for calculational purposes, we consider in this non perturbative expansion
the relation between the pole mass $M_{\rm pole}$\ and the running mass $M_{b}$\
in the appropriate order of the coupling constant \cite
{JAMINLANGE,CHETYRKINSTEINHAUSER1,VERMASEREN}.

Finally, on the left hand side of equation (\ref{SR}), we parametrize the
absorptive part of the two point correlation function by means of a narrow
width approximation of the lightest $B_{q}$ resonance plus the hadronic
continuum of the $b\overline{q}$ channel starting at $s_{cont}$, above the
resonance region: 
\begin{equation}
\frac{1}{\pi }\,\,\mathrm{Im}\Pi (s)=M_{B_{q}}^{4}\,f_{B_{q}}^2
\,\,\delta (s-M_{B_{q}}^{2})+\frac{1}{\pi }\mathrm{Im}\Pi ^{\rm cont.}\theta
(s-s_{\rm cont.})
\end{equation}
where $M_{B_{q}}$ and $f_{B_{q}}$ are respectively the mass and the decay
constants of the lowest lying pseudoscalar meson $B_{q}$.

Looking back to equation (\ref{SR}) and taking into account all the theoretical
parameters as well as the mass of the $B_{q}$-meson as our inputs in the
calculation, we see that the decay constant can be computed as far as we
had a good control of the hadronic continuum contribution of the
experimental side.

Since this is not the case, to cope with this problem we make an appropriate
choice of the polynomial ($P(s)$) in that equation. We take: 
\begin{equation}
P(s)=a_{0}+a_{1}s+a_{2}s^{2}+a_{3}s^{3}+\ldots +a_{n}s^{n},
\label{POLY} 
\end{equation}
such that its coefficients are fixed by imposing a
normalization condition at threshold $P(s_{\rm ph.}=M_{B_{q}}^{2})=1$, and
requiring that should vanish in the range $\left[ s_{\rm cont.},s_{0}\right] $ 
in a least square sense, i.e., 
\begin{equation}
\int_{s_{\rm cont.}}^{s_{0}}s^{k} \, P(s)\,\,ds=0\,\,\mathrm{for}\,\,k=0,\ldots n%
\label{POLY2}
\end{equation}
These conditions exactly cancel the continuum contribution as far as $%
\mathrm{Im}\,\,\Pi ^{\rm cont.}$ can be well approximated by an $n$ degree
polynomial. On the other hand, by virtue of the normalization condition,
it will enhance the role of the $B_{q}$
resonance. Notice however that by increasing the degree of the polynomial, $%
P(s) $, we will require the knowledge of further terms in the non
perturbative series, which are unknown. Therefore, a compromise criterion on
the choice of the polynomial degree has to be taken.

To check the consistency of the method in this work, we have considered a
second and third degree polynomials and the results are fully compatible within
the range of the errors introduced by the inputs of the calculation. We also
have checked explicitly that a smooth continuum contribution has no
influence in the result. This procedure was previously used in the
calculation of the charm mass from the $c\overline{c}$ experimental data.
The continuum data from the BES II collaboration \cite{bes} had no influence
when an appropriate polynomial was included \cite{PENARROCHASCHILCHER}.
Employing the same technique, very accurate prediction of the bottom quark
mass was also obtained using the experimental information of the upsilon
system \cite{PENARROCHABORDESSCHILCHER}.

After these considerations we proceed with the analytical calculation of the
decay constant $f_{B_{q}}$,\ neglecting the contribution that comes from $%
\mathrm{Im}\Pi ^{\rm cont.}$. For different powers of $s^{i}$ we obtain
different sum rules which, in principle, should give the same result for the
decay constant $f_{B_{q}}$. Notice however that, by increasing the value of $i$
we get larger contribution from both, the large $s$ region of the spectral function and higher orders in 
the non perturbative QCD expansion\footnote{The results obtained with higher moments are compatible with the
ones found here. However they are less stable 
with the duality parameter $s_0$ and, hence, less accurate.}.
Since these pieces have the main experimental and theoretical uncertainties, we will
consider the calculation
coming from the sum rules corresponding to the powers with $i=0$ and $i=1$ which, in any case it is
enough for our purposes. 
\begin{equation}
M_{B_{q}}^{4} \, f_{B_{q}}^{2}=\frac{1}{\pi }%
\int_{\left( M_{b}+m_{q}\right) ^{2}}^{s_{0}}P(s)\mathrm{Im}\Pi ^{\rm pert.}(s) \, ds \, -
{\rm Res} \left\{ P(s)\Pi ^{\rm nonpert.}(s), \, \, s \, = \, M_{b}^{2}\right\} 
\label{FIRSTSUMRULE}
\end{equation}
for $i=0$ ("first sum rule") and 
\begin{equation}
M_{B_{q}}^{6} \, f_{B_{q}}^{2}=\frac{1}{\pi }%
\int_{\left( M_{b}+m_{q}\right) ^{2}}^{s_{0}}sP(s)\mathrm{Im}\Pi ^{\rm pert.} (s)ds-
{\rm Res} \left\{ s \, P(s) \, \Pi ^{\rm nonpert.} (s), \,\, s \, = \, M_{b}^{2}\right\} 
\label{SECONDSUMRULE}
\end{equation}
for $i=1$ ("second sum rule"). Let us emphasize that in these two sum rules
there are two unknowns, the decay constant $f_{B_{q}}$ and the QCD duality
threshold $s_{0}$, the last one appears in the upper integration limit and
also in the coefficients of the polynomial $P(s)$. Therefore, we can use
both sum rules to determine $f_{B_{q}}$ as well as $s_{0}$. To employ a
couple of sum rules in order to fix the QCD duality threshold ($s_0$) is a usual
procedure \cite{REINDERS,JAMINLANGE}, but, as we will see, with our method we have
the additional advantage that the value obtained for $s_{0}$ is very stable,
in the sense that any change of this value would not affect appreciably the
result of the decay constant $f_{B_{q}}.$

\section{Results}

In the calculation of $f_{B}$ we take $m_{q}=0$ in the factor $%
(M_{b}+m_{q})^{2}$ of the correlation function, and in the low integration
limit of equation (\ref{EQU}), whereas for  $f_{B_{s}}$ we keep the strange quark mass different from
zero although its contribution turns out to be negligible.

The experimental inputs are as follows. For $q$ being the light quarks $q=u, \, d$
we have the physical threshold $s_{\rm ph.},$ at the squared mass of the lowest
lying resonance in the $b\overline{q}$ channel :

\begin{equation}
s_{\rm ph.}=M_{B}^{2}=5.279^{2}\,\,\mathrm{GeV}^{2}.  \label{BU}
\end{equation}
The continuum threshold $s_{\rm cont.}$ is taken at the $B\pi \pi $ \
intermediate state, in an $I=\frac{1}{2}$ s-wave, i. e. 
\begin{equation}
s_{\rm cont.}=\left( M_{B}+2m_{\pi }\right) ^{2}=30.90 
\mathrm{GeV}^{2}\,. 
\label{SCONTB}
\end{equation}

For the strange $s$ quark we take: 
\begin{eqnarray}
s_{\rm ph.} & = & M_{B_{s}}^{2}=5.369^{2}\,\,\mathrm{GeV}^{2} \nonumber \\ 
s_{\rm cont.} & = & \left( M_{B_{s}}+2m_{\pi }\right) ^{2}=31.92\,\,\mathrm{GeV}^{2}.
\label{BS}
\end{eqnarray}

In the theoretical side of the sum rule the inputs we take are as follows.
The strong coupling constant is taken at the scale of the electroweak $Z$
boson mass \cite{BETHKE} 
\begin{equation}
\alpha _{s}(M_{Z})=0.118\pm 0.003  \label{ALPHA}
\end{equation}
and run down to the computation scale using the four loop formulas of
reference \cite{SANTAMARIA}. The values, with the corresponding error bars,
for the quark and gluon condensates (see for example \cite{JAMINLANGE}) and
the mass of the strange quark \cite{PDG} are: 
\begin{eqnarray}
\langle \overline{q}q \rangle (2\,\mathrm{GeV})\,
&=&(-267\,\pm \,17\,\,\mathrm{MeV})^{3}, \nonumber \\
\langle \frac{\alpha _{s}}{\pi} \,G\,G \rangle &=&0.024\,\pm \,0.012\,\,\mathrm{GeV},\nonumber  \\
\langle \overline{q}\sigma G q\rangle \, & = &m_{0}^{2} \, \langle \overline{q}
q \rangle,\,\,\,\left( \mathrm{with}\,\,\,m_{0}^{2}\,=\,0.8\,\pm 0.2\,
\mathrm{GeV}\right)\nonumber  \\
\langle \overline{s}s\rangle &=&(0.8\pm 0.3)\langle \overline{q}q\rangle\nonumber  \\
m_{s}\left( 2 \, \mathrm{GeV}\right) &=&120\,\pm 50\,\mathrm{MeV}
\label{INPUT}
\end{eqnarray}
We also need to fix the renormalization scale $\mu $ in the equations
(\ref{PERT}) and (\ref{NOPERT}). For definiteness we take $\mu
=M_{b}(M_{b})$, although in order to find the stability of the results under
the renormalization scale we will study its variation for values of this scale 
in a range appropriate for our discussion, namely, ($\Lambda_{QCD}^{2}\,\ll \,\mu ^{2}<\,s_{0})$.

Finally for the bottom quark, the value $M_{b}\approx 4.20\,\mathrm{GeV}$ is
nowadays generally accepted. We take the result of 
\cite{PENARROCHABORDESSCHILCHER} which has also been obtained with the sum rule
method described here: $M_{b}(M_{b})\,=\,4.19\,\pm 0.05\,\mathrm{GeV}$.

Now, we proceed in the way described before. Firstly, we compute$
f_{B}$ as a function of $s_{0}$ with the two different sum rules (\ref
{SR}), for $i=0,1$. Then, we fix $s_{0}$ at the value where both sum rules
give the same result for $ f_{B}$. With this usual procedure, and for a
second degree polynomial, we find $s_{0}=48.5 \,\mathrm{MeV}%
^{2}\,\ $ and $\,\ f_{B}=\,185\,\mathrm{MeV}$.

\begin{figure}[th]
\centerline{\psfig{figure=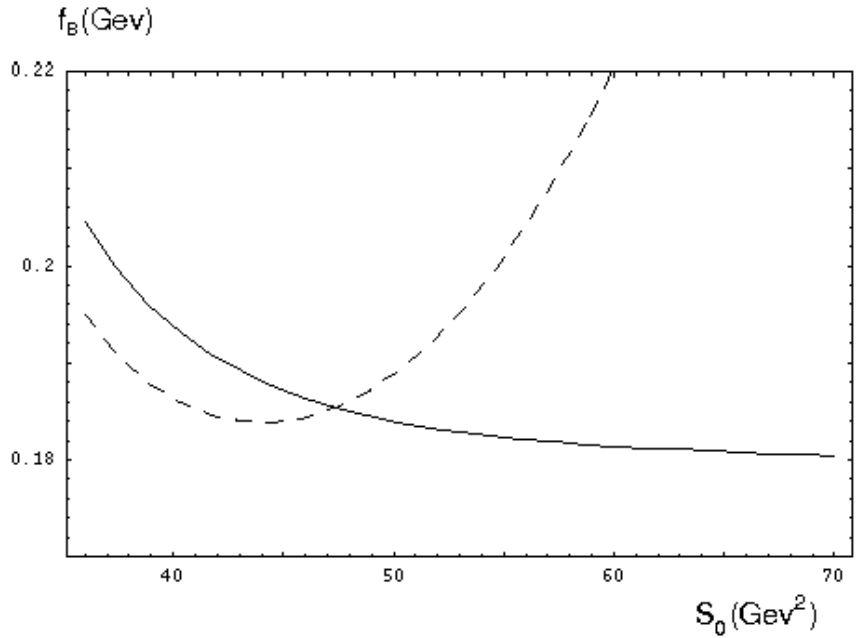,scale=0.7}} 
\caption{Decay constant $f_{B}$ as a function of the integration radius 
$s_{0}$ for $M_{b}(M_{b})=4.19\,\mathrm{GeV}$. With a second degree polynomial in
the sum rule (\ref{SR}), the dashed line represents the case $i=1$ and the solid line the case $i=0$.}
\end{figure}

From the theoretical inputs quoted above the main source of errors comes
from the bottom mass which, in the range given above, produces a variation
in the decay constant of $\mp \,11\,\mathrm{MeV}$. Other source of errors are
the quark condensates\footnote{As for the contribution of
the condensates, the only relevant comes from the lowest dimension one
which gives an $8\%$ of the total result. The higher dimension terms considered here
doesn't give a sizeble contribution giving a hint on the convergence of the condensate series
in this approach.} which affect the result by 
$\pm 4\,\mathrm{MeV}$. 
Since the decay constants are physical observables, the results should be independent
of the renormalization scale $\mu$. However, here we have used an approximation in
the asymptotic spectral function of QCD, taken only the two-loop order (\ref{PERT},\ref{NOPERT}).
Therefore a residual dependence on the renormalization scale is expected. In order
to quantify the uncertainty of fixing the scale at the bottom mass, we vary the
scale in the range $\mu \,\in \,\left[ 3,6\right]
\,{\rm GeV}$, introducing an uncertainty of $\pm 18\,{\rm MeV}$ in the result. This dependence
in the renormalization scale is expected to lower down if higher orders in the 
coupling constant in (\ref{PERT},\ref{NOPERT}) are taken into account.

Adding quadratically all these errors, we finally quote the following result
for the decay constant of the light meson $B$: 
\begin{equation}
f_{B}\,=\,185\,\pm \,22\,\ \mathrm{MeV}.  \label{RESULTFB}
\end{equation}

Notice in figure 1 that with $i=1$ (dashed line) and a second degree polynomial there is a stable value
(minimum), $f_{B}=\,183\,\mathrm{MeV,}$ at $s_{0}=44\,{\rm GeV}^{2}$. But for $%
i=0$ (solid line) the result $f_{B}=\,180\,\mathrm{MeV,}$ is an stable
value (inflexion point) which is practically constant around $s_{0}=60\,$\textrm{GeV}$^{2}$ 
However, to compare with
other results from QCD sum rules, we take the crossing point of figure
1 as our final result. The stable results of both curves are completely compatible
within error bars in such a way that we could have considered the value for $f_B$ in
the wide range of the stability region of $s_0$.

Proceeding in the same fashion for the $B_{s}$ meson, with the only change
given by the nonzero mass of the light $s$ quark, we find the decay constant
$f_{B_{s}}$. In this case, as can be appreciated in figure 2, the $s_{0}$
value where the two sum rules give the same result for $f_{B_{s}}$ is $%
s_{0}\,=\,49.6$ $\mathrm{GeV}^{2}$.

\begin{figure}[th]
\centerline{\psfig{figure=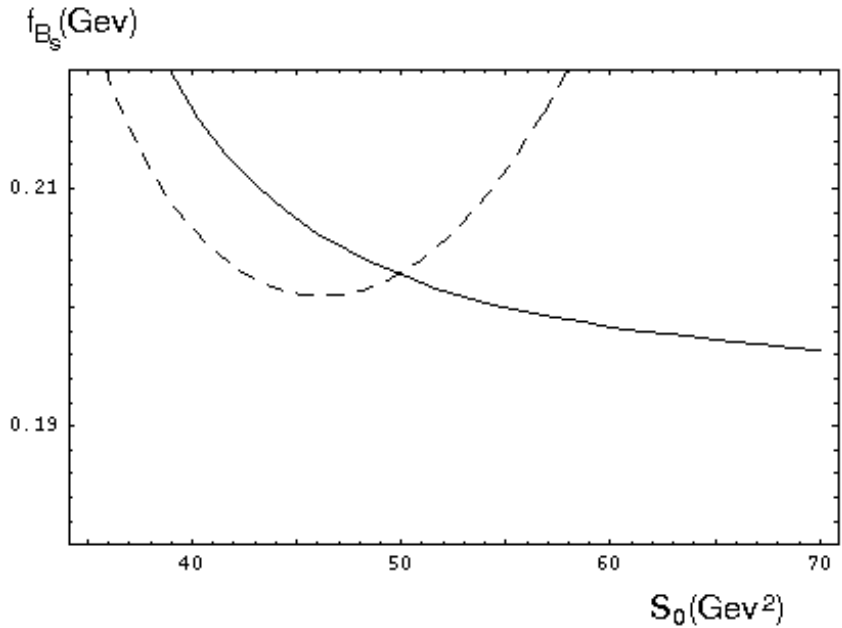,scale=0.77}} 
\caption{Decay constant $f_{B_{s}}$ as a function of the integration radius 
$s_{0}$ for $M_{b}(M_{b})=4.19\,\mathrm{GeV}$. With a second degree polynomial in
the sum rule (\ref{SR}), the dashed line represents the case $i=1$ and the solid line the case $i=0$.}
\end{figure}
The result in the intersection point is: 
\begin{equation}
f_{B_{s}}\,=\,202\,\pm \,24\,\,\mathrm{MeV}.
\label{RESULTFBS}
\end{equation}
where a similar analysis of errors has been considered. The only new
ingredient is the uncertainty coming from the strange quark mass, which
contribution turns out to be negligible.

It is of special interest the ratio of the decay constants $f_{B_{s}}$ and $%
\ f_{B},$ which should be $1$ in the chiral limit. We find
\begin{equation}
\frac{f_{B_{s}}}{f_{B}}\,=\,1.09\,\pm \,0.01
\end{equation}

We are free to remark that in the calculation of this ratio, the uncertainties of the theoretical
parameters are correlated, this is why the final error becomes very small.

\bigskip

\section{Conclusions}

In this note we have computed the decay constant of $B_{q}$-mesons for $q$
either the strange $s$ or the $u$ or $d$ massless quarks. We have used a
suitable combination of moments of QCD finite energy sum rules in order to
minimize the shortcomings of the available experimental data. On the
theoretical side of the sum rule, we have used the pseudoscalar two point
function calculated up to two-loop in perturbative QCD and up to dimension
six condensates in the non perturbative QCD expansion. Instead of the
commonly adopted pole mass of the bottom quark, we use the running mass to
get a good convergence of the perturbative series. We have a good control of
the results against the duality threshold $s_{0}$, which turn to be very
stable.

The results found, taking the running mass of the bottom quark $%
M_{b}(M_{b})\,=\,4.19\pm 0.05\,\mathrm{GeV}$ are given in equations (\ref{RESULTFB},\ref{RESULTFBS})
and we collect here for convenience: 
\[
f_{B}\,=\,185\,\pm 22\,\mathrm{MeV},
\,\,\,\,\,\,\, f_{B_{s}}\,=\,202\,\pm 24\,\mathrm{MeV}.
\]
where the error bars come from the uncertainty in the theoretical parameters (%
\ref{INPUT}) as well as the residual dependence on the
renormalization scale. We notice that the results are very sensitive to the value of
the running mass, giving most of the theoretical uncertainty. On the other
hand they turn out to be quite stable against the variations of the other input
parameters, in particular the integration radius $s_{0}$.

In this treatment we could not include the three-loop corrections in the two point
correlation function, since the complete analytical QCD expression along the
cut is not known. Despite of this lack of information, one can interpolate
the three-loop low energy QCD calculations with the high energy ones \cite{JAMINLANGE,CHETYRKINSTEINHAUSER}.
The results do not differ much from the exact two-loop calculation. 
Another possibility, is to include the
three-loop high energy expansion  along the integration circle $\left| s\right| =s_{0}$
which is known to certain accuracy.
We have also performed this calculation somewhere else, having results very
close to the ones found here.

Within the error bars, our results agree with other results in the
literature, obtained with different sum rule methods and with lattice
computations \cite{REINDERS,JAMINLANGE,NARISON,DOMINGUEZPAVER,APE,UKQCD}.
However, we advocate for values of the decay constants in the lower band.

\end{document}